# A meta-analysis of the mechanical properties of ice-templated ceramics and metals

Sylvain Deville[1*], Sylvain Meille[2], Jordi Seuba[1]

[1] Laboratoire de Synthèse et Fonctionnalisation des Céramiques, UMR3080 CNRS/Saint-Gobain, 84306 Cavaillon, France.

[2] Université de Lyon, INSA-Lyon, MATEIS CNRS UMR5510, F-69621 Villeurbanne, France

* Corresponding author. Email: sylvain.deville@saint-gobain.com, Twitter: @DevilleSy

Ice templating, also known as freeze casting, is a popular shaping route for macroporous materials. Over the past 15 years, it has been widely applied to various classes of materials, and in particular ceramics. Many formulation and process parameters, often interdependent, affect the outcome. It is thus difficult to understand the various relationships between these parameters from isolated studies where only a few of these parameters have been investigated. We report here the results of a meta analysis of the structural and mechanical properties of ice templated materials from an exhaustive collection of records. We use these results to identify which parameters are the most critical to control the structure and properties, and to derive guidelines to optimize the mechanical response of ice templated materials. We hope these results will be a helpful guide to anyone interested in such materials.

## Introduction

Ice templating, or freeze casting[1], has become a popular shaping route for all kinds of macroporous materials. The process is based on the segregation of matter (particles or solute) by growing crystals in a suspension or solution (Fig. 1). After complete solidification, the solvent crystals are removed by sublimation. The porosity obtained is thus an almost direct replica of the solvent crystals.

Ice templating has been applied to all classes of materials, but particularly ceramics over the past 15 years. Although a few review papers [1–6] have been published, they mostly focus on the underlying principles. Little can be found on the range of properties that could be achieved.

Ice templating is a complex process, where many parameters can affect the results: formulation of the suspension (nature of the solvent, particle size of the starting powder, solid loading, binder, surfactant, nature of the material, pH, viscosity), freezing conditions (setup, temperature, cooling rate), and measurements conditions (sample dimensions, loading rate, setup). Every article within the scientific literature provides a set of these parameters. The influence of only a few of these parameters has been systematically investigated. The complex interdependent relationships of these parameters cannot be captured in a single, isolated paper.

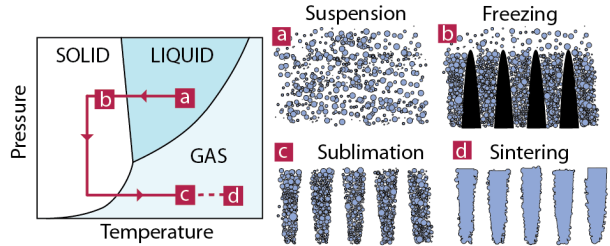

*Figure 1: Principles of ice-templating. The colloidal suspension is frozen, the solvent crystals are then sublimated, and the resulting green body sintered.*

The many names used for this process makes the literature review more difficult. Throughout the paper, we refer to the process as *ice templating*, probably the most popular term, albeit not the most appropriate. *Ice templating* implies that water is used. A variety of solvents have been used to date: water[8], camphene[9], tert-butyl alcohol[10], naphthalene/camphor[11], dioxane[12], terpene[13], deep eutectic solvents[14], or carbon dioxide[15]. Various terms have been coined for the process: ice templating[16], freeze-casting[1], crystal-templating[17], directional freeze-drying[18], freeze-thaw[19] freeze gelation[20], ice-segregation-induced self-assembly[21], or thermally induced phase separation[22]. *Frozen solvent templating* [23] is probably the most generic.

Here, we performed a meta-analysis of the published structural and mechanical properties of ice-templated materials, to both provide an objective overview of the range of properties that can be achieved and to identify the critical parameters that control these properties. The combination of data and parameters from a large collection of papers [8–12,24–142] can partially make up for the incomplete records in the literature. We use the results from this meta-analysis to derive guidelines to optimize the structure and mechanical response of ice templated materials.

## Methods

The meta-analysis was performed by manually data mining published, peer-reviewed papers or conference proceedings for various parameters dealing with the structural and the mechanical properties as well as the formulation of the suspensions and the characteristics of the powders. The papers analyzed cover a variety of materials including ceramics (alumina[8–11,26,28,32,36,38,43,48,54,58,77,81,89,102,104,105,111,112,131,143], zirconia[49,80,100,101,103,125–127], silicon carbide[45,106,130], silicon nitride[35,69,72,75,76,79,84,87,88,128], mullite[90,110,124,137], hydroxyapatite[12,37,44,52,61,68,78,86,91,95–98,107,109,113,115,120,134,135], tri-

calcium phosphate[31,40,56,117,118], lead zirconate titanate (PZT)[39,63,74,82,83,121,122], barium titanate[73], silica/kaolinite[47], titania[24,29,50,99], zirconium diboride[51], chromium carbide[42], sialon[64,66], lithium iron phosphate[60], yttrium orthosilicate[57,62], cermets[116,123], cordierite[129], aluminium nitride[133], iron oxide[136], baghdadite[34]), glass and bioglass[41,55,85,114,117,117,132], and metals (titanium[65,70,92–94,108]). The plots were digitized when the exact values were not provided directly in the text or through tables. The data were analyzed and plotted using an IPython notebook [144]. The raw data (CSV files) as well as the IPython notebook are available from FigShare (http://dx.doi.org/10.6084/m9.figshare.1412626).

The pore size is not always reported. In ice templating, crystals are usually grown unidirectionally. A morphological gradient that corresponds to the initial transient regime (nucleation of the crystals and growth to reach the steady growth state) is systematically observed[145], but its dimensionsare usually limited to a few hundred microns at most. It is therefore not representative of the overall microstructure, and often removed before testing. We focus here of the characteristics of the pores that correspond to the steady state growth regime. The resulting pore size is thus determined by a measurement of the cross-section of the crystals in a direction perpendicular to the main solidification direction (given by the temperature gradient).

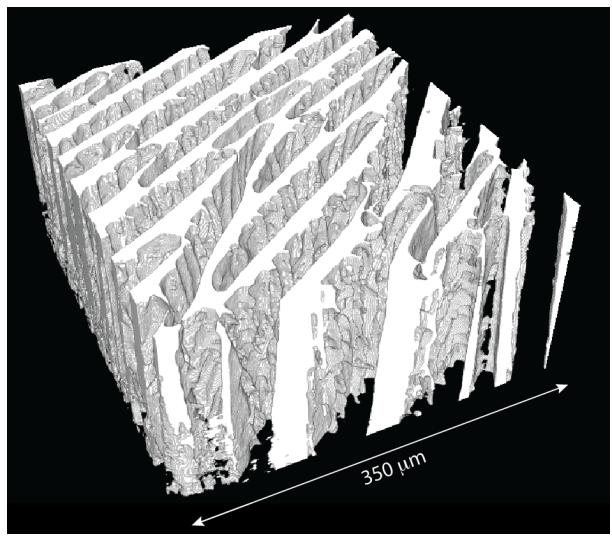

*Figure 2: Three-dimensional reconstruction of the structure of a macroporous ice templated alumina sample (water-based suspension), obtained by X-Rays computed tomography. Ice crystals were grown in the vertical direction.*

The pore morphology strongly depends on the nature of the solvent. When water is used, the pores usually have a lamellar morphology or, more precisely, the pores are tubular with an ellipsoidal cross-section, and a strong aspect ratio between the short and long axis (Fig. 2). The *width* of the pore, which corresponds to the short axis of the ellipse, is usually reported. If the pores exhibit an isotropic cross-section, the pore size is given directly by the dimension of the pore in its cross-section. Since the morphology of the pores is usually maintained along the solidification direction, there is usually no need to correct the measured pore size by stereological parameters[146]. In addition, the pore size is often reported as a single number; the actual pore size distribution is seldom reported, whereas is might have a critical influence over the mechanical response.

The presence or absence of structural defects was determined by a visual inspection of the micrographs. As the most commonly encountered structural defects (described below) are usually oriented perpendicular to the main solidification direction, micrographs taken along the solidification direction are required to ascertain the presence or absence of defects. It was thus not always possible to conclude on their presence.

# Results

We first discuss the reported structural properties (porosity, pore size). Then, the compressive and flexural strength values are discussed, and the influence of processing and measurement parameters over these strength values is reported and analyzed.

## Structure

For a complete description of the morphologies and typical structural features found in ice templated materials, we refer the reader to one of the reviews already published on the process [1–5,7]. Here, we focus on the amount of porosity and the pore size, which impact the reported strength values.

The porosity is almost always measured after sintering, which reduces the pore size and content by 10 to 25 vol.% [147]. The porosity is therefore not approaching 100% as the solid loading gets close to 0 vol.%. Achieving very high porosity is easier with materials that do not require sintering (polymers, graphene, nanotubes).

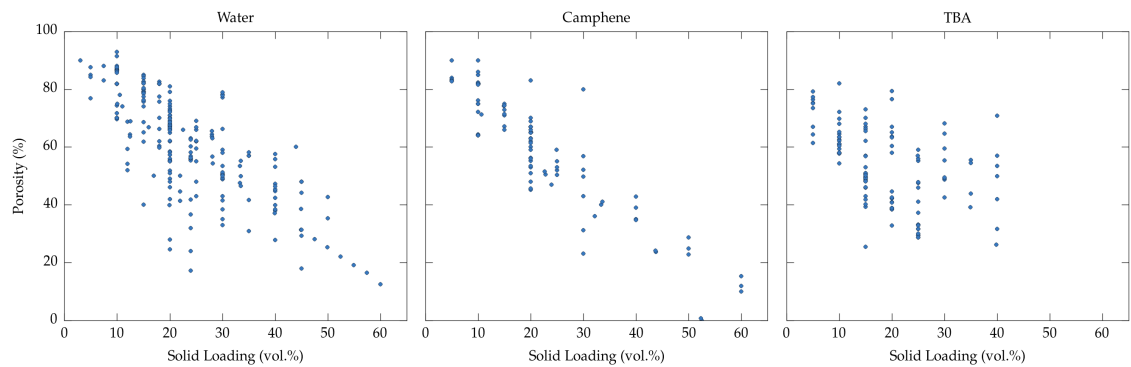

*Figure 3: Porosity vs. solid loading, for the three most commonly used solvents: water, camphene, and TBA. The porosity reported here is the total porosity of the macroporous materials after sintering.*

The total porosity is controlled, as a first order parameter, by the initial solid loading of the suspension. Usually, the growing crystals repel all particles, additives, and impurities, and concentrate them within the inter-crystal space. The final porosity is a replica of the solvent crystals: its total content can thus be adjusted directly by the solid loading. We plotted, in figure 3, the total porosity vs. the solid loading of the suspension for the three most commonly used solvents: water, camphene, and tert-butyl alcohol (TBA).

Ice templating is a versatile and flexible process, able to achieve a very wide range of porosity, from extremely high porosity (>99.9 vol.% for ice-templated graphene[148,149] for which no sintering is necessary) to almost no porosity after sintering[48,133]. A transition in the morphology of the pores is observed when the solid loading is so high that the redistribution of particles by the growing crystals cannot take place anymore. The typical solid loading for this transition is around 50-55 vol.%[1]. Above this critical solid loading, the low tortuosity of the individual pore channels is lost, and the pore size becomes smaller.

The range of pore size achieved is plotted in figure 4, for water, camphene, and TBA. The typical pore size range is essentially the same for all the solvents, from 4 to 500 μm. Pores larger than 200 μm were not found with TBA, although this limit might just be a consequence of the lower number of records.

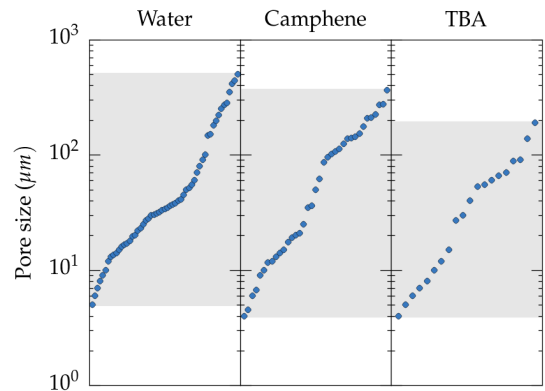

*Figure 4: Range of pore size achieved for the three most commonly used solvents: water, camphene, and TBA.*

A unique capability of ice-templating is the ability to yield directional pores with an almost constant cross-section. There is nevertheless almost always a progressive increase of the pore size as we move along the solidification direction. This increase results from the direct relationship between the growth velocity of the crystals and the pore size, and the difficulty to maintain a constant growth velocity over long distances (centimeters). A constant growth velocity can be achieved[150] with two cold sources at each extremity of the mold and proper cooling rate profiles but has not been used very often. The main parameter that controls the pore size is the growth velocity, which results from the cooling rate and the applied temperature gradient[151]. It is therefore strongly dependent on the experimental setup used, and cannot be ascertained from this meta-analysis, as almost every paper reports a different cooling setup. The relationship between the cooling rate and interface velocity has nevertheless been already explored in detail in individual papers[45].

The relationship between the pore size and the total porosity is shown in figure 5. As expected, larger pores can be achieved for greater porosity. Figure 5 and 6, in concert, also illustrate that a trade-off between pore size, porosity, and mechanical properties can be achieved for a porosity just above 55 vol.%. The achievable pore size increases rapidly when the porosity exceeds 55 vol.%. Further increase of the porosity results in moderate increase of maximum pore size that can be achieved, while the strength values rapidly falls, as shown and described below (Fig. 6).

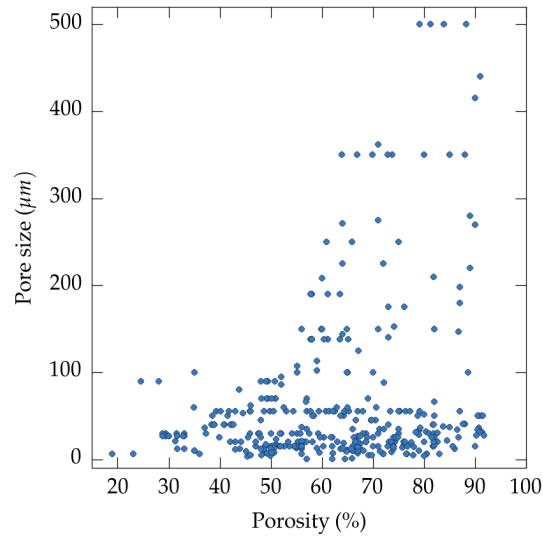

Figure 5: Pore size vs. total porosity, all solvents. Large pores (>100 μm) can rapidly be obtained when the porosity exceeds 55 vol.%. Small pores (<30 μm) can be obtained for all solid loadings; a fast growth velocity of the crystals is enough to achieve small dimensions.

## Overview of the mechanical properties: compressive and flexural strength

Two main properties of ice templated materials have been reported in the literature: the compressive strength and the flexural strength. Ice templated materials are almost always anisotropic: hence the compressive strength, along the most favorable direction, is usually reported. The flexural strength values of ice templated materials were only reported in a handful of papers.

**Compressive strength** To identify the critical parameters independently of the nature of the materials used, the strength values have been normalized by the flexural strength of the corresponding dense materials. This normalization can be questioned, in particular for low porosity where the fracture mode is not primarily bending. For cellular ceramic materials, the local stresses on the solid phase is in majority bending[152]. The structures are always imperfect, failure occurs thus always in local bending[153]. This normalization allows a comparison of all types of materials, from very strong (zirconia, alumina, titanium), to very weak (barium titanate, calcium phosphates, etc.) materials.

We plotted the relative compressive strength vs. porosity in figure 6, for all solvents. To provide a comparison with well-known morphologies, data for cellular ceramic materials[154] have been plotted on the same figure. These data had also been normalized by the flexural strength, and provide a fair overview of the mechanical properties of a wide range of ceramics materials (alumina, silicon carbide, silicon oxycarbide, titania, hydroxyapatite, and mullite) and shaping routes (polymer replica, wood replica, sacrificial templating, direct foaming with surfactants or particles). To a first approximation, there is nothing specific about the properties of directional, ice templated materials. Even if the experimental dispersion is relatively important, the compressive strength can be predicted from the porosity within an order of magnitude. There are more data points at high porosity for ice templated materials, as the specific set of papers selected by Studart *et al.* was focused on cellular ceramics where a high porosity is usually targeted.

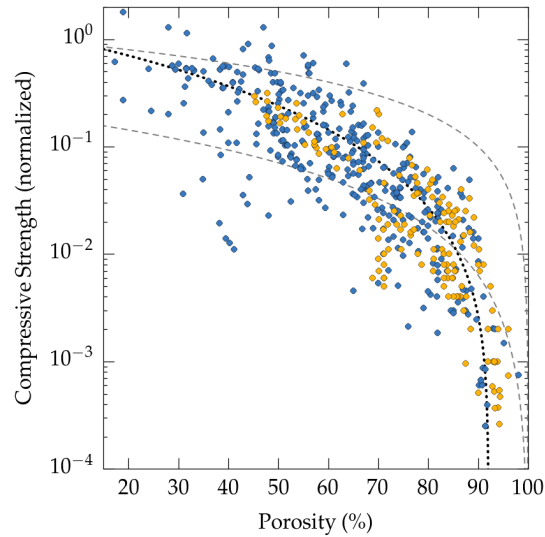

*Figure 6: Overview of the compressive strength achieved (all materials and all solvents), plotted versus porosity. Color code: blue: ice templated materials, yellow: cellular ceramics from reference [154]. The upper and lower dashed lines correspond respectively to the close cell and open cells models. The black dotted line corresponds to the best fit of a regression analysis with a power law (exponent=2.26).*

**Flexural strength** Little flexural strength data of ice templated materials have been reported. The values, reported in figure 7, have again been normalized by the flexural strength value of the corresponding dense materials. Since less data can be found than for the compressive strength, it is thus more difficult to extract any trend from these values.

Very few data points can be found for high porosity (>70%). This absence is probably related to the difficulty of machining samples with parallel faces, and the risk of indentation by the support during the loading. The comparison of figures 6 and 7 shown that the strength values in flexion and compression are very similar, for a given porosity. The porosity is therefore the first order parameter to test. There is therefore little interest to test these materials in flexion.

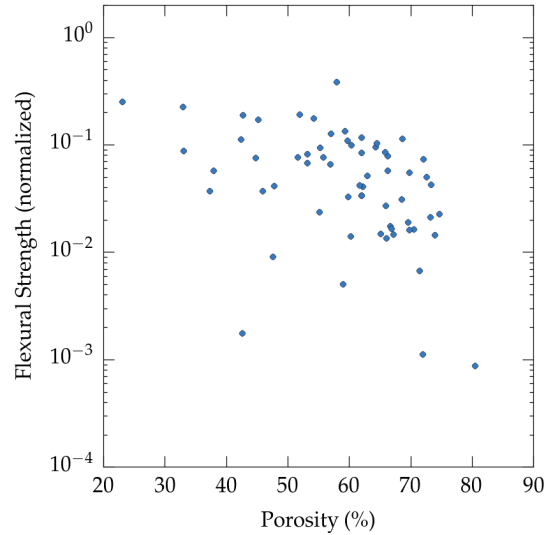

*Figure 7: Overview of the flexural strength achieved (all materials, all solvents). Data from references [10,26,27,32,35,47,66,69,72,75,76,79,84,87,88,135]*

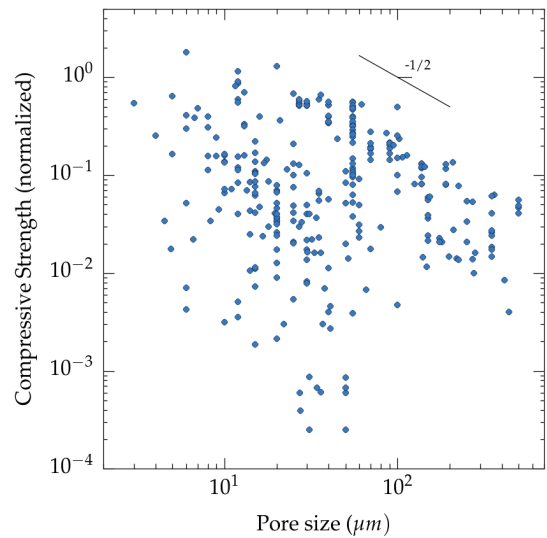

*Figure 8: Compressive strength vs. pore size (all materials, all solvents). Data from references [26–28,30–47,49,51–54,56–62,64–68,71,73,77–79,85,86,90,91,94–96,98,100,103–109,118,119,125–127,129,130,132,134,135]*

**Pore size**. The pore size seems to influence the strength of ice-templated materials (Fig. 8). There is again a lot of scatter of the values, resulting from variations in the formulation and process parameters, but the maximum strength that can be achieved is inversely proportional to the pore size. When the pore size is smaller than a few tens of μm (approx. 20 μm), little increase in strength is apparently found, although this is most likely just due to the paucity of data in that region. For a given pore size (e.g. 10 μm), the strength values are scattered over several orders of magnitude. It is thus clear that many other parameters have a critical influence over the mechanical properties of the ice templated materials.

**Failure mode** Two failure modes can be observed for porous ceramics tested in uniaxial compression: brittle or cellular [155]. A damageable, cellular-like behavior, is observed at high porosity (typically above 60 vol.%), with a progressive densification of the material after local bending of the solid walls, while for porosity below 60 vol.%, a brittle behavior is observed with long macro-cracks propagation along the loading direction. With ice templated materials, the transition between the two failure modes is observed around 50 vol.% (Fig. 9). The stress-strain curves of the mechanical tests are not reported very often, and the majority of the ice templated materials have a porosity above 50vol.%. Many features such as the pore size, pore morphology, or density of the walls, can impact the local loading mode of the structure and its mechanical response. Therefore, we do not expect a very sharp transition between the two failure modes when such a large number of papers is analyzed. The precise determination of the transition between the failure modes would thus require additional investigations.

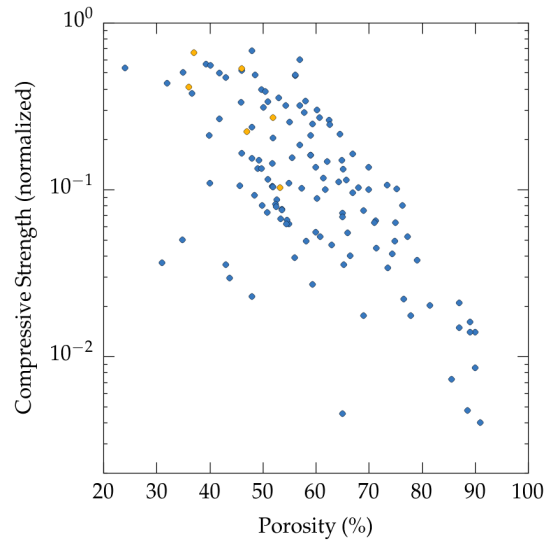

*Figure 9: Compressive strength and failure modes (all materials, all solvents): brittle (yellow)[25,64,85] or cellular (blue)[25,26,28,33,40,41,51,58,64,65,70,85,90,91,94–96,98,103,127,132,135].*

For isotropic porous ceramics, this transition is governed by the size of the crack initiating from the large pores and the mean distance between these pores. Here, the pore morphology is different, but the amount of stored energy favors brittle fracture at high densities. On the contrary, for the lowest densities, the low amount of stored energy and the presence of defects in the microstructure favor the progressive damage of the material during compression.

**Structural defects**

The presence or absence of structural defects is critical to determine the overall mechanical response of the porous materials. Defects in ice templated materials are obtained under certain conditions, which, as of today, have not been clarified[156]. These defects are related to the development of so-called *ice lenses*: ice crystals growing perpendicular to the solidification direction, which adopt a crack-like morphology (Fig. 10). Their appearance is mostly studied in geophysics, for their occurrence is of interest to explain, in particular, the freeze-thaw cycles of frozen soils[157–159].

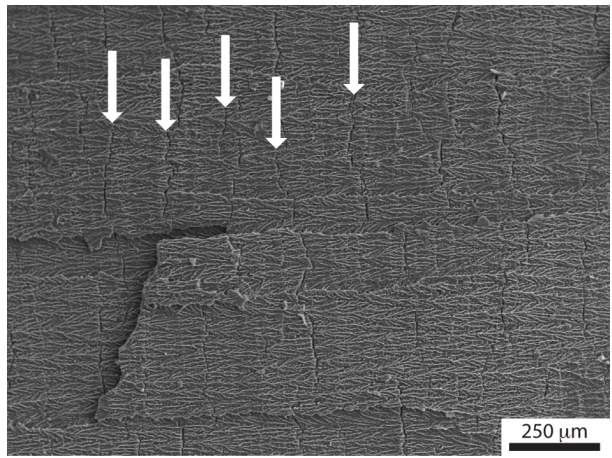

*Figure 10: Ice-lens type defects (arrows) in ice-templated alumina. Solidification direction: left to right. Such defects appear perpendicular to the main solidification direction.*

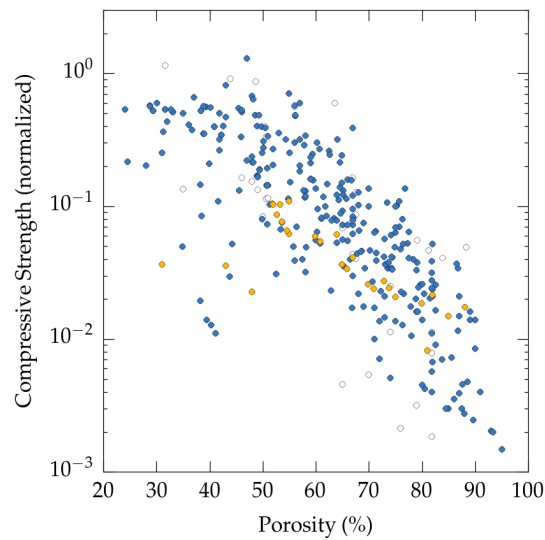

*Figure 11: Compressive strength in presence and absence of ice-lenses type structural defects (all materials, all solvents). Color code: blue: no defects [26–31,33,36,38,40,41,43,44,46,50-55,61,64,65,67,68,70,71,73,77,78,85,90,91,94,98,100,103–110,113,115,118,124–127,129,130,132,134,135,143], yellow: defects [25,52,86,91,96], white: cannot tell from the published micrographs [37,39,49,56,91,95,96,119].*

The most appealing hypothesis[160] invoked to explain their formation is related to the development of residual stresses during the completion of freezing (a result of the volume increase that accompanies the water to ice phase transformation), and mechanical resistance of the pack of concentrated particles between the ice crystals. The morphology of the defects is indeed very similar to mechanically induced cracks[160].

It is nevertheless clear that:

- Ice-lens type defects have only been observed when water is used. This may be explained by the negligible volume change upon liquid to solid phase transformation of all the solvents used but water.

- The presence of such defects is extremely damageing for the strength of the materials, as shown in figure 11. When such defects are found, the materials is essentially pre-cracked and the resulting strength, at a given porosity, is always low compared to a defect-free (structural macro-defect) ice-templated material. Avoiding such defects is nevertheless a necessary but not sufficient condition to obtain high strength values [156], as some defect-free ice-templated materials have a low compressive strength, compared to ice-templated materials with the same porosity.

## Suspensions properties

Several parameters are set when the initial suspension to be ice templated is prepared: nature of the solvent, solid loading, addition of processing additives such as binder or surfactants, etc. A wide variety of processing additives have been used, it is thus not possible to extract any conclusion from this meta-analysis with regards to their possible influence. Two main parameters have been investigated here: the nature of the solvent and the particle size of the starting powder.

**Particle size** The influence of the particle size over the compressive strength (for all solvents) is plotted in figure 12. Similar to the previous plots, a lot of scatter is observed, and the most informative is to look at the upper limit of what can be achieved. There seems to be an optimal range of particle size (approx. 0.5-2 µm) for which the compressive strength is maximized. This optimum may have several explanations. When large particles are utilized as raw materials for this processing, gravitational sedimentation of raw materials in a slurry can always occur during casting or freezing. Thus, for porous metals, the freeze casting has been applied mostly for titanium, because of its light density. Since the spacing between crystals is mostly given by the temperature gradient, particle size must be always smaller than the cell wall thickness. It the particle size is larger than the average spacing between crystals, a good templating effect by the crystals cannot be achieved. Large particles are thus not very favorable.

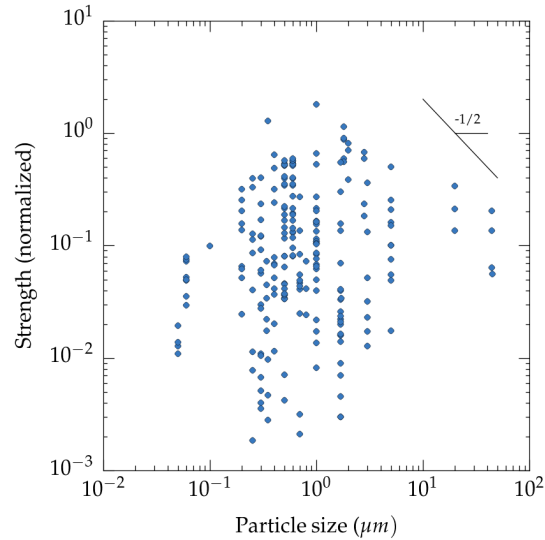

*Figure 12: Influence of the starting particle size over the compressive strength (all materials, all solvents). Data from references [26–28,31,36–39,46,49,51–53,56,61,64,65,67,68,70,71,73,77,78,85,94–96,103–106,113,118,119,125–127,129,130,132]*

In addition, the particle size dictates, along with the sintering conditions, the grain size in the final ceramic (or metallic) material. In ceramic and metal materials, the strength increases when the grain size decreases[161]. For ice templated materials, an increase of the particle size and hence the grain size would thus decrease the compressive strength. In addition, the grain size also plays a role on the characteristics of inter granular pores obtained after sintering. The larger the particle size, the more difficult the densification of the walls surrounding the pores. Hence, when we use larger particles (a few µm's or more), we are more likely to end up with inter granular pores, which will act as defects when the material is put under stress/strain. The resulting strength is thus lowered.

The decrease of compressive strength when the particle size becomes lower than approximately 0.5 µm is more difficult to explain. A possible explanation could be the role of the inter particle electrostatic forces, since ceramic particles usually exhibit a surface charge at their surface. When the particle size decreases, the magnitude of the electrostatic interactions increases. This increase may in turn affect the behavior of the particles when they are concentrated by the growing crystals, and affect the packing and thus the homogeneity of the structure by an increase of the inter-particle porosity. Again, additional experiments would be required to investigate this aspect.

**Role of the solvent**

The choice of the solvent is critical for the structure and properties obtained. Since the pores are a replica of the solvent crystals, both pore morphology and pore size are dependent on the nature of the solvent. In figure 13, we plotted the normalized compressive strength for the three most commonly used solvents. The size of the markers is proportional to the pore size. Several conclusions can be drawn from these plots.

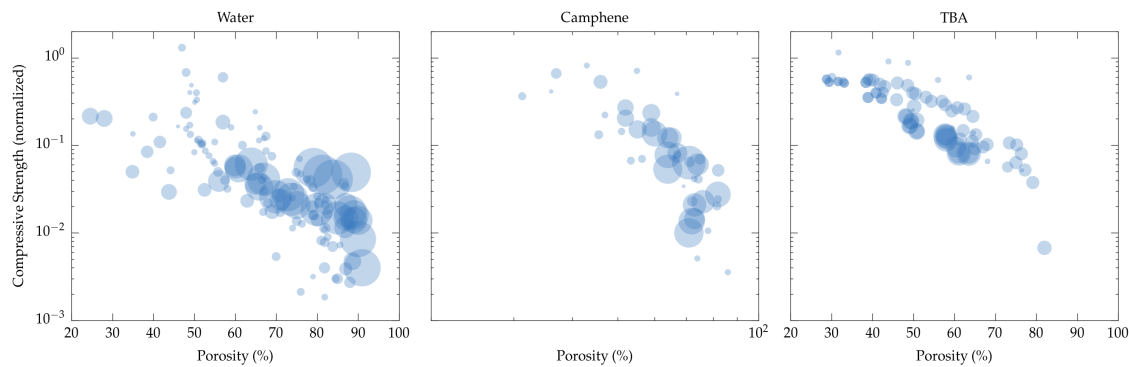

*Figure 13: Influence of the nature of the solvent over the compressive strength. References for the data: water[8,12,26,27,31,33–38,43–49,51,52,55–61,68,69,72–76,81,84,86–92,94–101,103,104,111,113,116,119,124,128–132,134,135], camphene[9,29,30,42,50,64–67,70,80,85,93,102,105–109,114,115,117,118,120–123,125–127,136], TBA[10,32,39–41,53,62,63,71,77,78,82,83,110,137]. The marker size is proportional to the pore size.*

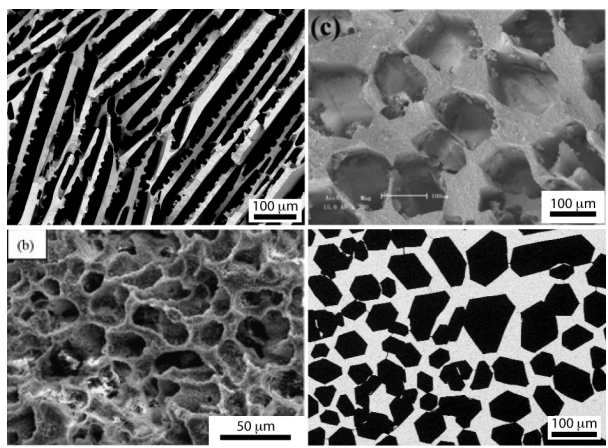

*Figure 14: Porous structures representative of what can be achieved with (a) water, (b) TBA, (c) camphene, and (d) water with additives (zirconium acetate). Credits: (b) Copyright (2010) Elsevier, used with permission from [162], (c) Copyright (2010) Elsevier, used with permission from [125].*

Different behaviors are observed for each solvent. With water, a lot of scatter is observed. Conversely, data with camphene and TBA show less scatter, which cannot be explained by the range of pore size that can be achieved, since it is essentially the same for all solvents (Fig. 4). Beyond the ice-lens defects discussed previously, the scatter with water can be partly explained by the wider range of pore morphology that can be achieved with water-based suspensions. Ice growth is very sensitive to the nature and characteristics of common ceramic processing additives (dispersant, binder, etc.)[163]. Morphologies typical of that obtained with several solvents are shown in figure 14. The growth behavior of camphene and TBA crystals is apparently a lot more robust. The crystals morphology is less sensitive to the formulation and process parameters. The pore size also provides useful observations. For high porosity (above 60 vol.%), large pores yield higher compressive strength values, but only when water is used. This increase possibly results from the greater thickness of the walls between the pores, which are therefore less sensitive to structural defects. For lower porosity (below 60 vol.%), smaller pores result in higher compressive strength values. For camphene and TBA, this is not evident, possibly because of the relative paucity of data for these solvents.

## Measurement conditions

The conditions under which the mechanical properties are measured are also critical, and determine, to a large extent, the absolute values obtained. Even though standard test methods and good practices exist, they are not systematically followed. The strain rate, the sample volume, the sample aspect ratio, the experimental setups, the surface preparation: all of these parameters affect, to some extent, the measured compressive strength. More subtle parameters, such as the presence of a pad (typically cardboard) between the sample and the testing setup to ensure a good alignment of the sample along the loading direction, are very rarely reported although they may also affect the results. We analyzed the influence of the strain rate, sample volume, and sample aspect ratio. There is too much scatter (due, in part, to all the parameters described in the previous sections) in the values reported here, so no clear trend could be observed with the current meta-analysis. Some of these parameters have been investigated in previous papers (see for instance the effect of loading rate reported in figure 6 of Fu *et al.*[96]).

## Discussion

## Prediction by the mechanical models

Several models have been developed, in particular by Gibson and Ashby[153,164], to predict the mechanical response of cellular solids with various morphologies (Fig. 15). The open- and closed-cell foam models, in particular, are very often used to understand and predict the behavior of cellular materials. These models predict the following values for the compressive strength.

- Closed-cell

$$\sigma = \sigma_p \left( C_6 \left( \phi \frac{\rho^*}{\rho_s} \right) + C_6''(1-\phi) \frac{\rho^*}{\rho_s} \right) \text{ (eq. 1)}$$

with $C_6 = 0.65$ and $C_6''=1$, and where $\phi$ is the solid fraction in the edges, $\rho_s$ and $\rho^*$ the apparent density of the dense and cellular material respectively, and $\sigma_p$ the modulus of rupture of the solid cell wall material.

- Open-cell (brittle crushing mode)

$$\sigma = \sigma_p C_4 \left( \frac{\rho^*}{\rho_s} \right)^{3/2} \text{ (eq. 2)}$$

with $C_4 = 0.2$, and where $\sigma_p$ is the modulus of rupture of the solid cell wall material, and $\rho_s$ and $\rho^*$ the apparent density of the dense and cellular material respectively.

- Honeycomb, out of plane

$$\sigma = 6E_s \left(\frac{\rho^*}{\rho_s}\right)^3 \quad (eq.\,3)$$

where $E_s$ the Young's modulus of the corresponding dense material, and $\rho_s$ and $\rho^*$ the apparent density of the dense and cellular material respectively.

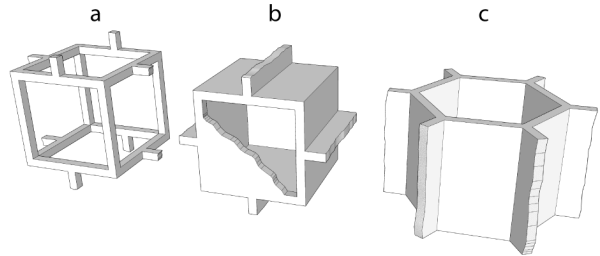

*Figure 15: Morphology of the foam cells of the Ashby models: (a) open-cell, (b) closed-cell, (c) honeycomb.*

As shown on figure 6, the first two models do not provide a good description of the behavior of ice template materials. The variation of the compressive strength with porosity does not follow the prediction of models for open- and closed-cell foams. Those models are well fitted for the prediction of mechanical properties of ceramic foams, because the cell walls in the foams are not continuous (they are divided by pores, similar to that of open cell model).

The morphology of ice templated materials —continuous pores, with an almost constant cross-section— is actually a lot closer to that of a honeycomb (Fig. 14). We plotted in figure 16 the measured compressive strength versus the compressive strength predicted by the honeycomb out-of-plane model[165] for water, camphene, and TBA. Although these three solvents yield crystals with very different morphologies, the high pore directionality makes the honeycomb model a good descriptor of the structure and indeed the observed strength.

These three solvents yield crystals with very different morphologies.

The camphene crystals (Fig. 14c) have a dendritic morphology[166]. That makes them less similar to the ideal honeycomb structure. This effect causes a disparity between the predicted and experimental values (Fig. 14b), particularly at high porosities where the growth of secondary arms (dendrites) is more important.

TBA crystals (Fig. 14b), on the other hand, are hexagonal and usually continuous along the solidification direction [53,78,162]. The agreement between the measured and predicted properties in this case is surprisingly good. This behavior may probably be extended to water crystals, when they adopt hexagonal morphologies (Fig. 14d) in the presence of zirconium acetate [167].

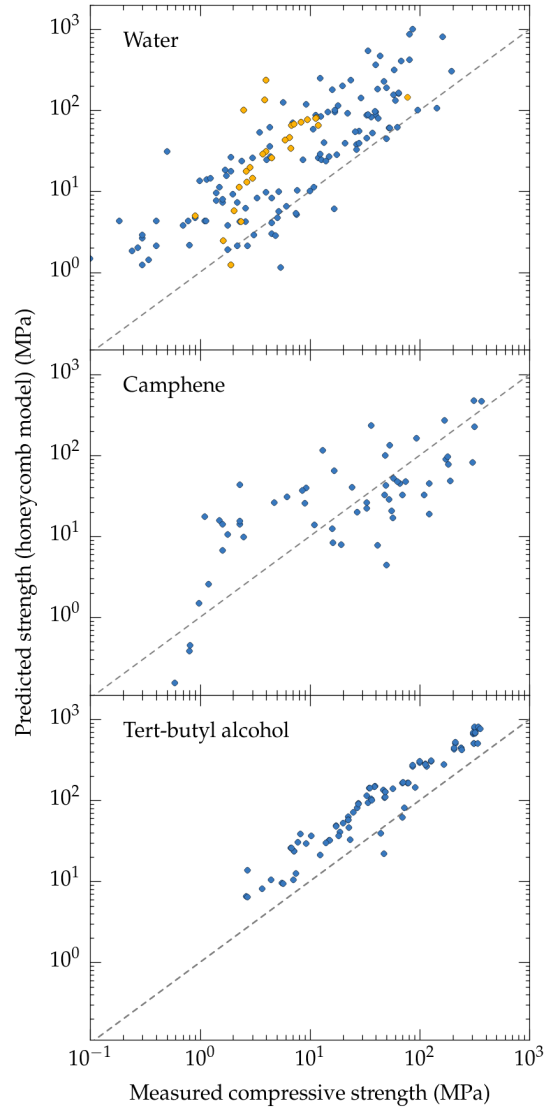

*Figure 16: Compressive strength: measured (all materials) vs. predicted by Ashby honeycomb out-of-plane model. The dashed line corresponds to the predictions of the model. Color code: blue, no defects; yellow: ice-lens type defects.*

The case of water deserves special attention. Even though the lamellar morphology with some degree of surface roughness does not resemble the honeycomb, the predictions from honeycomb are in reasonable agreement with the experimental values (Fig. 14a), mainly because the predominant solicitation in both cases is Euler buckling [168]. Nevertheless, there is still a lot of scatter in the predictions. This scatter may come from two different sources. First, although it was observed the same failure mechanism, the stress distribution in the struts might be different for both morphologies. Then, as mentioned previously, ice crystals are more sensitive to the addition of additives, which may affect the directionality of pores and the generation of structural defects.

In addition, in the use of the honeycomb model, the density of the "dense" material is not corrected for wall porosity, as might be the case for large particle sizes. This is difficult to estimate without a measurement of the porosity in the wall, which is almost never mentioned or reported. This might account for some of the scatter in the prediction.

## Missing data and the difficulty of systematic comparison

The current meta analysis underlines again the crucial importance of reporting all the process and experimental parameters, and, whenever possible, a complete structural and mechanical characterization. We believe a meta-analysis provides an objective view of what we may achieve in terms of structure and properties. Nevertheless, such analysis, and any other systematic comparison, is more easily performed and more reliable if all parameters and properties are carefully documented in every paper. Unfortunately, each author has its own habits for reporting results. The solid loading of the suspension, for instance, can be reported either in weight or volume percent, and computed with or without the other processing additives (binder, dispersant). The materials are often not fully characterized, and key informations are often missing. Systematic comparisons are therefore more difficult.

## Which conditions are the best ones ?

This, of course, depends on what one wants to achieve in terms of structural or functional properties. The following lessons can be learned from the current meta analysis:

- The targeted range of pore size should not dictate the choice of the solvent, as all of the most popular solvents (water, camphene, TBA) can fulfill a very similar range of pore size (1-500 µm). The morphology of the pores can nevertheless be a criterion that dictates the choice of the solvent, in particular if certain functional properties are targeted. The dendrites aligned along the growth direction of ice crystals, for instance, can be beneficial [169] for neurite growth.

- There is apparently an optimal range of particle size (approximately 0.5-2 µm) for which the compressive strength is maximal. The reason for this optimum is not clear.

- It is less robust to work with water-based suspensions than with camphene or TBA. The growth of ice crystals is more sensitive to impurities, formulation, and process parameters. Using water therefore requires a good understanding of the process and the various parameters involved.

- The open- and closed-cell foam mechanical models are not very appropriate to predict the behavior of these cellular materials. The only exception is the honeycomb model of Zhang and Ashby, which accurately predicts the behaviors of TBA crystals-templated samples.

- The influence of the pore size on the mechanical response is complex. For high porosity of water-based samples, larger pores are more likely to yield a high compressive strength. For low porosity (all solvents), smaller pores are preferable.

- A trade-off between large pore size and mechanical properties can be achieved for a porosity above 55 vol.%. Such trade-off is often critical in applications like tissue engineering, where large pores are required and yet the materials must be able to sustain a certain level of load. The achievable pore size increases rapidly when the porosity exceeds 55vol.%. Further increase of the porosity results in a moderate increase of the pore size, while the strength rapidly decreases (Fig. 6). To achieve both large pores (>200 μm) and high strength (50% of the strength of the corresponding dense material), the optimal porosity is therefore in the 60-65 vol.% range. It is very likely that this trade-off can be further optimized towards higher strength by a careful process and structural control.

A lot of work remains to be done to fully understand the behavior and the range of properties that can be achieved with ice-templated materials. The following points probably deserve more attention in future work:

- The Weibull modulus (which requires a very large number of samples), has just been reported in one study [170] for such materials (zeolite), and yet is critical to predict the reliability of these materials, in particular for structural applications.

- The detailed identification of the transition between the brittle and cellular failure modes could be determined more precisely. This understanding will be helpful to select the proper mechanical response based on the functional requirements of the targeted application.

- These materials are almost always anisotropic. The direction that provides the best mechanical response is almost always tested; a lot could be learned from testing the materials in their weak direction. Fu *et al.*[96], for instance, reported a strength ratio of 3 between strong and weak directions.

- The role of the pore size distribution over the mechanical response has not been investigated, and might play a critical role for the mechanical response and failure behavior. To have a better understanding of the influence of the solid phase (respectively the porosity) morphology, on the mechanical properties, a quantitative study has to be carried out. The use of 3D volumes from X-Rays computed tomography, coupled with a finite elements simulation [171] or with DEM modeling [172] are powerful approaches.

## Conclusions and recommendations

Ice templating is a rich processing route, with many formulation and process parameters, which can be used with practically any material. This versatility nevertheless results in added complexity. It is essential to have a good understanding of the process and the influence of the formulation, freezing, and testing parameters. We hope that this meta-analysis will be a helpful guide to anyone interested in working with such materials in order to optimize their relative mechanical strength.

We also insist that it is essential to report properly all the experimental parameters and results, and in a way that can be easily reused by anyone else. It is, for instance, much easier to extract values from tables than from plots. Until data repositories become standard and data mining can be automated, it is therefore a good practice to include both tables and plots.

## Acknowledgments

The research leading to these results has received funding from the European Research Council under the European Union's Seventh Framework Programme (FP7/2007-2013) / ERC grant agreement 278004 (project FreeCo). JS has been supported by the CNRS and Saint-Gobain under BDI grant agreement 084877 for the Institut National de Chimie (INC).